# A Perspective on the Grouping and Categorization of Nanomaterials

Scott C. Brown

The Chemours Company, Newark, USA

Email: scott.c.brown@chemours.com

**Abstract**

The development of science-based categorization strategies for regulatory purposes is not a simple task. It requires understanding the needs and capacity of a wide variety of stakeholders and should consider the potential risks and unintended consequences. For an evolving science area, such as nanotechnologies, the overall uncertainties of designing an effective categorization scheme can be significant. Future nanomaterials may be far more complex and may integrate far different functionalities than today's nanomaterials. There is much that has been learned from our experience with legacy nanomaterials and particulate substances in general. Most of today's nanomaterials are not new nor dramatically different from parent or existing chemical substances, however there are some nuances. Applying these learnings to define reasonable science-based categories that consider how different emerging nanomaterials might be from existing known substances (while integrating sound concepts as they develop) would be a pragmatic and flexible path forward. However, there are many barriers down this road including a need for improvement and updates to chemical classification systems to improve hazard and risk communications, while promoting transparency and consistency.

**Introduction**

According to the Organization for Economic Cooperation and Development [1], "*a chemical category is a group of chemicals whose physicochemical and human health and/or ecotoxicological properties and/or environmental fate properties are likely to be similar or follow regular patterns, usually as a result of structural similarity*". One of the benefits of the use of chemical categories is that closely related chemicals are considered as a group, or a category, rather than as individual chemicals… [accordingly] *"not every chemical needs to be tested for every endpoint. Instead, the overall data for that category should prove adequate to support a hazard assessment… [and] must enable an estimate of hazard for the untested endpoints"* [1]. Data gap filling in a chemical category can be filled out by several methods including read-across, trend analysis, and (external) (Quantitative) Structure Activity Relationships ([Q]SARs). Technical guidance documents are available from the OECD regarding how to apply this [1]. In 2013, the OECD Council, through a legal instrument, recommended that, "*its Member Countries apply existing international and national chemical regulatory frameworks to manage the risks associated with manufactured nanomaterials*" [2]. It was noted that these frameworks and existing test guidelines may need to be adapted to consider specific properties of nanomaterials. The OECD had been working since 2006 to develop approaches for risk assessment for manufactured nanomaterials that are of high quality, science-based, and internationally harmonized. Collectively, this implies that the information on nanomaterial hazards can be derived from the respective bulk material, from molecules or ions of its constituents, or from similar nanomaterials.

There are a range of grouping approaches that have been suggested for nanomaterials. Many of which integrate the knowledge of the "parent" chemical substance [3-8], and others rely more explicitly on exposure potential [9-11], modes of action [12], integrated approaches for testing and assessment (13-17), and alternative screening tests [18, 19]. A detailed review of different nanomaterial categorization strategies has been published as part of a World Health Organization initiative [20].

There are many drivers for the categorization of nanomaterials. For industry, it provides an opportunity to reduce animal testing, while (ideally) helping to draw a common understanding of key factors that drive safety or undesirable effects. The latter is especially important for product stewardship practices such as Responsible Care® [21] where product safety considerations are taken into account early on in the product development process. The DuPont and Environmental Defense Fund (EDF) NanoRisk Framework [22] developed over a decade ago, is an embodiment of the core principles behind Responsible Care® and is also in line with the concept of safer-by-design. For regulators, sound categorization principles would also clarify decision making and streamline approval and restriction processes while aiding to identify critical information gaps also helping with their mission. The public-at-large benefits from the safe use of nanotechnologies that continue to transform our everyday lives.

**Discussion**

While there are many benefits for initiating a categorization scheme for nanomaterials, there are also potential drawbacks or inefficiencies. Like any other form of categorization, if the strategy employed is not fit-for-purpose, then the process can become more of a barrier than a benefit. For nanotechnologies, this could mean significant barriers to innovation that would take away from the societal benefit of the technology, or it could mean that more harmful materials may be placed on the market resulting in a negative perception and deselection of the technology. It is clearly important that any chosen categorization strategy be grounded on sound science, and that periodic assessments be made that the chosen path remains fit-for-purpose rather than a matter of convention.

There are parallels between the categorization of nanomaterials for regulatory purposes and the categorization of products on the market. For both to be effective, there needs to be identified metrics and clear guidelines for ensuring comparable, consistent and fit-for-purpose data generation and information exchange. While this statement seems obvious, its implementation is a common stumbling block. If products are not appropriately tested (e.g., are not tested to for the right performance or quality criteria), there is little assurance that it would meet the needs of customers and a high risk that the material may be dissatisfactory, resulting in loss of continued sales. Similarly, using an inappropriate testing procedure could lead to inappropriate categorization and inappropriate safety assumptions from flawed read-across or other measures to fill data gaps. The lack of clarity in communications, absence of consistent standard procedures, and questionable data quality are unfortunately real issues in nanosafety research [23]. This has led to confusion and a disproportionate sense of risk associated with nanomaterials resulting in challenges to all stakeholders. In recent years, substantial steps have been taken to right the course through block-funded research initiatives, notably FutureNanoNeeds, NanoReg, ProSafe, CEINT, and others. However, additional efforts are clearly required and there are high hopes for more recent projects such as GRACIOUS as well as the essential efforts by the OECD and the International Organization for Standardization (ISO).

Any successful strategy must have:

1) Alignment and clarity on the purpose for the categories
2) Sufficient experience with the substances to be categorized (or similar substances) and the targeted hypothesis testing (an understanding of caveats)
3) An overarching meaningful (fit-for-purpose) assessment strategy including standard methods and media (where appropriate) with appropriate benchmarks and controls.
4) A common understanding of suitable data quality metrics, availability of comparable data and sufficient diversity in data.

However, the challenge is greater than this. There remain questions of whether the right parameters are being considered and whether those parameters (and associated testing) are valid for the range of nanomaterials on which they would be used. The characterization of many of the seemingly simple parameters have been met with difficulties, such as particle size distribution [24, 25] and surface chemistry. These parameters will only become more difficult to characterize with more complex $2^{nd}$, $3^{rd}$, and $4^{th}$ generation nanomaterials. Strategic foresight should be used to consider the range of probable impacts (including both positive and negative outcomes) to better identify, whether specified arbitrary cutoffs for categorization (e.g., 50% of particles having a minimum dimension below 100nm by number for nanomaterial status) are appropriate, and to what extent there is a need of explicit guidance on appropriate levels of characterization (e.g., for defining surface chemistry) and detailed reporting to ensure meaningful, comparable defensible, and helpful information on a broad spectrum of materials in the future.

Beginning in 2011, the European Commission (EC) embarked on a journey applying a threshold approach for identifying nanomaterials through the Recommended Definition of a Nanomaterial in 2011 [26]; currently under revision as of May 2020) and more recently implemented this into the REACH framework. The introduction of the concept of *nanoforms* for grouping purposes via adaptation of the REACH annexes [8] has led to further questions. A nanoform is a group within a chemical substance that conforms to the EC Recommended Definition of a Nanomaterial that have a common size, particle shape and surface chemistry. Individual nanoforms or collections of nanoforms (sets) can be further grouped as categories if they have suitable similar human health, ecotoxicological and environmental fate properties. By creating *nanoforms* the EC assumes that size, shape and surface chemistry will always be relevant for the most basic groups. Yet, what if another property—that is not substantially impacted by size, shape or surface chemistry—defines adverse effects, such as a minor impurity or the presence of defects? In such a scenario, the *nanoform* strategy may be over-restrictive. Having detailed characterization of all substances is important, however, given the evolving status of nanotechnology with the expectation of increasingly complex materials in the future, having fixed, unproven baseline assumptions in categorization strategies may not be effective. There should be a clear science-based purpose rather than an arbitrary default stance. As the REACH registration deadline has approached and passed, the framework has morphed into hypothesis-driven sets of nanoforms for registration purposes. The future will only tell whether or not this approach will result in fit-for-purpose classification schemes, but what it has done is to highlight issues in the supporting science as well as a lack of readiness of available standards and contract laboratories to help with meaningful and comparable data filling of designated parameters.

The implied assumption that the safety of a nanomaterial principally relies on simple, isolated parameters properties like particle size distribution, surface chemistry, and particle shape is likely inaccurate. Deviations in the biological response to nanomaterials are influenced by a complex interplay between multiple parameters—and those parameters may be different for each nanomaterial type—but may feed into a common behavior that influences the overall biological or environmental effect. Appropriately and consistently measuring these behaviors across a variety of nanomaterials such that meaningful comparisons can be drawn is a complicated task but far more efficient than attempting to measure all possible parameters that may contribute to the behavior [27]. Mechanistic testing for phenomenon or behaviors related to processes that feed into behaviors related to hazard or exposure concerns are likely to be more useful with targeted regulatory purpose in mind and associated quality criteria [28, 29]

For instance, particle dissolution behavior has been shown to be an important parameter related to the potential hazards for substances that have hazards associated with their constituent ions and is also an important factor related to the residence time of respirable particles as solid particles *in vivo*. Dissolution behavior describes the release of ions and biodurability of a material. The factors that influence dissolution can be quite diverse and are influenced by size, shape and surface chemistry [30]. For each particle type the key mechanisms promoting or preventing dissolution may be different. In many cases, the dissolution behavior may be reasonably in agreement with the parent substance, however, deviations are known to exist. For instance, the dissolution behavior for some particle types may primarily be defined by how well the surface is coated and the apparent insolubility of that coating, rather than simply the surface chemistry [31]. Another nanomaterial with the same size, shape and surface chemistry (i.e., of the same nanoform) may have a substantially different dissolution behavior simply because its surface is not fully encapsulated or hermetically sealed. Measuring dissolution behavior directly —though complicated—is a far easier task than embarking on a detailed research project to identify all the potential surface and particle characteristics (including coating integrity) that lead into that behavior. Binning materials based solely on chemistry and not considering nuances that may impact key behaviors has the potential to be misleading and may further limit innovation. In many cases, how the behavior is manifested may not be known or may be related to confidential business information (e.g., trade secrets). Furthermore, the processing and handling history of the material can also impact behaviors like dissolution. Not only do basic physicochemical parameters matter but also the history of particle systems prior to the exposure event. All of these elements need to be taken into account for safer-by-design considerations. Categorization strategies should not provide barriers but rather should encourage utilizing the full set of knowledge to make safer product development choices. While product stewardship is an integral part of product development processes the risks of regulatory acceptance do weigh into the consideration of options and unintended consequences should be considered.

The categorization of nanomaterials should not occur without a clear purpose. Alignment and adequate communication are critical to attaining quality data and providing context. It is also necessary for opening channels for useful feedback to facilitate strategic and scientific advances. If the overall aim is to reduce testing while enhancing understanding to improve decision making, then barriers should not be put in place that would detract from achieving these goals. In many cases, the uncertainty of the testing methods and data quality need to be taken into account as well as the usefulness of data gathering exercises without well-defined methods or reference substances. The historical experience with particulate substances needs to be considered when available.

While nanomaterials are by themselves interesting, it is vitally important that our collective knowledge and experience with the bulk particulate chemical substances be integrated into considerations for nanomaterials as context. There is a considerable body of work on legacy nanomaterials and particle substances (Figure 1). This knowledge should be used as a baseline and differences between known substances and anticipated behaviors from this rich dataset should be applied to identify any unique and novel attributes of nanomaterials. In many respects, this is similar to the approach taken by the United States Environmental Protection Agency in their Section 8A reporting rule [32]. Focusing on nanomaterials that behave differently than expected will advance the science faster than recreating the wheel of particle toxicology by focusing on nanomaterials that simply are not different [33]. This approach is also consistent with more recent international movements looking towards Advanced Materials and the possible health and safety considerations that may need to be resolved as science and technology continue to evolve.

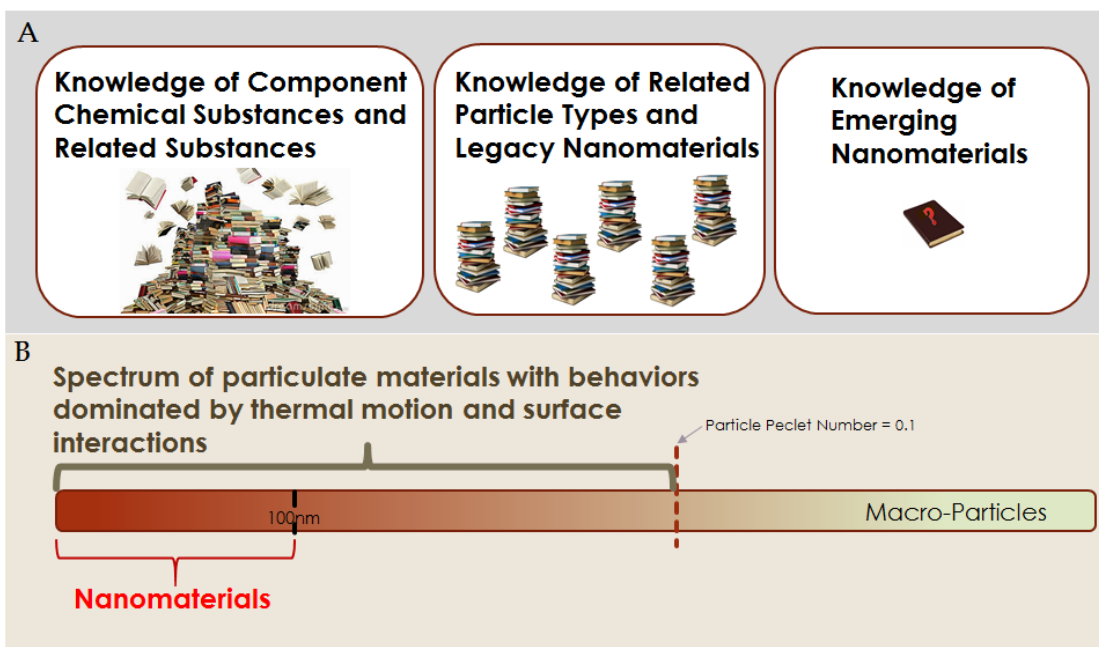

**Figure 1:** Nanomaterial Categorization schemes should leverage the bounties of existing knowledge and focus on similarities and differences with known substances. **(a)** Illustration of the relative knowledge difference between hypothetical component chemical substances and related substances, related particulate substances and legacy nanomaterials (1rst generation nanomaterials, e.g., carbon black, amorphous silica) and that for emerging nanomaterials. **(b)** A schematic illustrating that nanomaterials (in the absence of unique and novel properties) likely have a number of behavioral attributes similar to larger particle forms. Many larger particle systems also have similar dependencies on surface interactions (as described by thermal motion dominance in behavior by the particle Peclet number) and undergo many of the complicating processes observed for nanomaterials. Societies experience with those larger materials and their diversity should be taken into account when considering similarity requirements for forming categories.

Along these lines, it could be argued that the need for categories may increase with increasing intrinsic substance hazard or exposure risk. In terms of encouraging and facilitating safer-by-design and proactive product stewardship, activities having a sliding scale categorization requirement may be helpful (see Figure 2). For instance, if a low toxicity substance is coated by another substance, also of low toxicity and similar toxicological profile, then testing requirement should be not be as substantive and activities such as read across should not be as restrictive as for a higher toxicity material coated by the same low toxicity substance. In general, substances of considerable biological activity and significant exposure potentials should have more categories, requiring higher similarity requirements (because the risks are higher), before data is assumed. While this may initially be seen as creating more burden for high toxicity materials, it can also be seen as an opportunity, especially if those materials can be modified to reduce their hazard profile. A non-nanomaterial, but particulate, example of this is the coating of quartz to reduce or prevent pulmonary toxicity [34]. While benign coatings on low toxicity materials (e.g., $TiO_2$) tend not to have an effect on particle toxicity [35-38], the effect can be dramatic in the case of more highly toxic substances such as quartz. Likewise, if the potential for exposure could be reduced, then having additional categories to reflect this would also be useful and is likely to encourage safer nanomaterial design choices and related innovations where applicable. For substances where there is an existing rich knowledgebase and extensive historical experience, it is logical that the categories should reflect this knowledge; however, for new substances, additional precautions (also taking into account constituent chemical substances) and more stringent requirement might be required until a reasonable experience is obtained. For next generation and unique substances, similar consideration should come into play. The more unusual or complex a material is, the more caution one needs to take before assuming data for these substances.

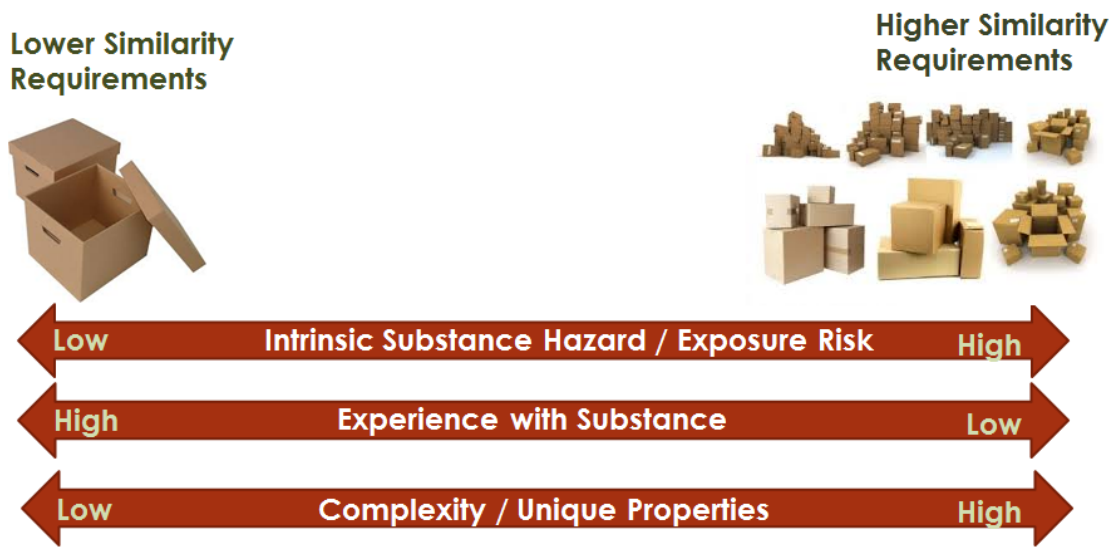

**Figure 2:** Schematic of the Right-Sized Categorization Strategy for Nanomaterials. Similarity requirements enabling grouping depends on the potential for hazard and exposure based on the composition, information from similar materials including the collective experience with those chemical substances, and how different the nanomaterial may be from substances that we already reasonably understand.

There are some barriers to making progress in the categorization of nanomaterials that are common to the chemical substance classification schemes, in general (see Figure 3). These include (1) the lack of an agreed upon regulatory-relevant, defined procedures and statistical metrics for proving a negative result, (2) the absence of potency considerations and too few categories for many of the hazard endpoints in the Globally Harmonized System for Hazard Communication (GHS), (3) the need for more rigorous, transparent, and consistent classification schemes that limit the use of arbitrary discretion in GHS, (4) consistent and transparent measures for assessing the quality and relevance of published non-guideline studies for regulatory purposes, and (5) integration of regional risk assessment policymakers with the regional GHS implementation process. Progress in fulfilling these elements would help bring clarity to a wide range of issues and stumbling points with the collective view of nanomaterials, emerging advanced materials as well as traditional chemical substances. Additional classification levels that reflect potency in a number of hazard categories would dramatically help with public communications of risk while also aiding in the assignment of similarity requirement in sliding scale, or right-sized categorization approach. It is clear that public perception in today's connected society has as much if not more impact on substance usage as regulatory science. Ensuring appropriate science-based communications requires more than "expert opinion" guidance but rather additional structure, rigor, and evaluation transparency. Consistency in interpretation of results including more transparent guidelines for "expert judgement" and weight of evidence considerations are essential stepping stones towards clarifying chemical and materials safety for the future.

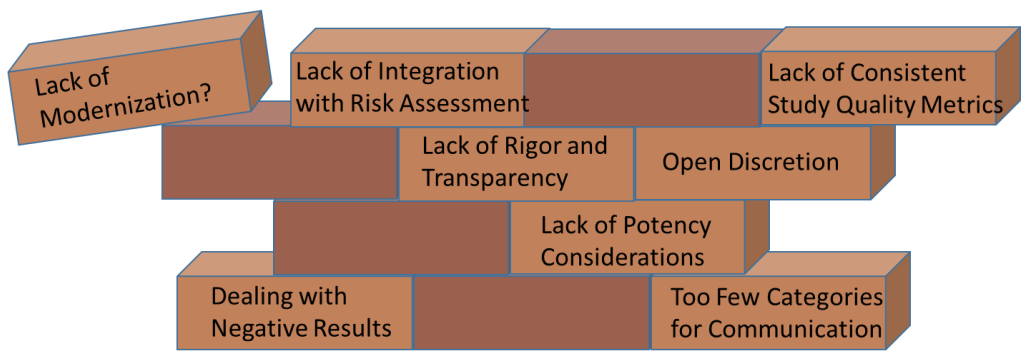

**Figure 3:** Barriers to categorization of nanomaterials that are common with issues periodically raised with modern chemical substance classification schemes.

While GHS is a landmark accomplishment, it is becoming increasingly apparent that the extent of harmonization can vary, and there are opportunities for further progress to be made. While this does not directly impact the choice of nanomaterial categorization approach, the hazard categories in GHS for component chemical substances could serve an important aspect of nanomaterial categorization. To aid in facilitating this streamlining, communication amongst risk assessors and the GHS expert community is essential. In many jurisdictions (including the US and Canada), the regional regulatory bodies focused on the risk assessment of chemical substances (including nanomaterials) and the authorities that implement and influence the development and improvement of the GHS system are different and often have distinct perspectives and priorities. By closing gaps and fortifying the robustness, consistency and transparency of chemical substance classifications, developing categorization schemes for both nanomaterials and chemical substances, in general, could be improved and better informed. Doing so may also make safer and more responsible choices more transparent for a broad range of manufacturers.

**Conclusions**

The design of effective nanomaterial categorization strategies should leverage existing knowledge and known consideration for particle substances and legacy nanomaterials. Categories and groups should not be based on a prescribed set of properties but derived from collective science and experience with dealing with the bulk substance(s), their hazard profile, the potential for exposure, and the existence of any unique properties for the nanoscale substance. Data gaps, data quality issues and related uncertainties for nanomaterials need to be considered and resolved. While categorization strategies should be beneficial, there are several concerns that develop if it is done without considering unintended consequences. Efforts should be made to facilitate strategies that would promote the use of safer-by-design principles and facilitate Responsible Care ® efforts in industry.

**Acknowledgements**

This article is one of a collection of articles about the categorization of nanomaterials, generated by research and workshop discussions under the FutureNanoNeeds project funded by the European Union Seventh Framework Programme (Grant Agreement No 604602). For an overview and references to other articles in this collection, see *The Nature of Complexity in the Biology of the Engineered Nanoscale Using Categorization as a Tool for Intelligent Development* by Kenneth A. Dawson.

The author is an employee of a company that manufactures, processes and uses particles and nanomaterials. Any opinions, findings and conclusions or recommendations expressed in this material are those of the author and do not necessarily reflect the views of the Chemours Company.

10. Gebel T et al (2014) Manufactured nanomaterials: categorization and approaches to hazard assessment. Arch Toxicol 88:2191–2211.
11. Bos PM et al (2015) The MARINA risk assessment strategy: a flexible strategy for efficient information collection and risk assessment of nanomaterials. Int J Environ Res Publ Health 12:15007–15021.
12. Kuempel, E.D., Castranova, V., Geraci, C.L., Schulte, P.A. (2012) Development of risk-based nanomaterial groups for occupational exposure control. Journal of Nanoparticle Research, 14(9), p.1029.
13. Oomen AG, Bos PMJ, Fernandes TF, Hund-Rinke K, Boraschi D, Byrne HJ, Aschberger K, Gottardo S, van der Kammer F, Kühnel D, Hristozov D, Marcomini A, Migliore L, Scott-Fordsmand J, Wick P, Landsiedel R (2014) Concern-driven integrated approaches to nanomaterial testing and assessment – Report of the NanoSafety Cluster Working Group 10. Nanotoxicology 8:334–348.
14. Stone V, Pozzi-Mucelli S, Tran L, Aschberger K, Sabella S, Vogel UB, Poland C, Balharry D, Fernandes T, Gottardo S, Hankin S, Hartl M, Hartmann N, Hristozov D, Hund-Rinke K, Johnston H, Marcomini A, Panzer O, Roncato D, Saber AT, Wallin H, Scott-Fordsmann, JJ (2013) Research Prioritization to Deliver an Intelligent Testing Strategy for the Human and Environmental Safety of Nanomaterials ITS-NANO Consortium, 128 p. Available at: http://www.itsnano.eu/wp-content/uploads/2013/12/ITS-NANO.pdf.
15. Arts JH et al (2015). A decision-making framework for the grouping and testing of nanomaterials (DF4nanoGrouping). Regul Toxicol Pharmacol 71(Suppl):S1-S27.
16. Braakhuis HM, Oomen AG, Cassee FR (2016) Grouping nanomaterials to predict their potential to induce pulmonary inflammation. Toxicol Appl Pharmacol 229:3-7.
17. Dekkers S et al (2016) Towards a Nanospecific Approach for Risk Assessment. Regul Toxicol Pharmacol 80:46-59.
18. Nel AE, Xia T, Meng H, Wang X, Lin S, Ji Z, Zhang H (2013) Nanomaterial toxicity testing in the 21st century: use of a predictive toxicological approach and high-throughput screening. Acc Chem Res 46: 607–621.
19. Lai DY (2012) Toward toxicity testing of nanomaterials in the 21st century: a paradigm for moving forward. Wiley Interdiscip Rev Nanomed Nanobiotechnol 4:1–15.
20. Landvik NE, Skaug V, Mohr B, Verbeek J, Zienoldiny S (2018) Criteria for grouping of manufactured nanomaterials to facilitate hazard and risk assessment, a systematic review of expert opinions. Regul Toxicol Pharmacol 95:270-279.
21. ACC (2012) Responsible Care® Product Safety Code of Management Practices. American Chemistry Council. https://responsiblecare.americanchemistry.com/Product-Safety-Code/
22. DuPont EDF (2007) Nano Risk Framework: A framework developed for the development of nanoscale materials. http://www.nanoriskframework.com/files/2011/11/6496_Nano-Risk-Framework.pdf
23. Krug HF (2014) Nanosafety research—are we on the right track? Angew Chem Int Ed 53: 12304-12319.
24. Linsinger T, Roebben G, Gilliand D, Calzolai L, Rossi F, Gibson P, Klein C (2012) Requirements on measurements for the implementation of the European Commission definition of the term 'nanomaterial'. Publications Office of the European Union. JRC73260. http://publications.jrc.ec.europa.eu/repository/handle/JRC73260.
10